\documentclass[%
 reprint,
superscriptaddress,
groupedaddress,
showpacs,preprintnumbers,
nofootinbib,
nobibnotes,
 amsmath,amssymb,
prb,
]{revtex4-1}

\usepackage{graphicx}% Include figure files
\usepackage{dcolumn}% Align table columns on decimal point
\usepackage{bm}% bold math
\usepackage{amssymb}   % for math
\usepackage{amsmath}
\usepackage{color}
\usepackage[normalem]{ulem}
\usepackage{siunitx}
\usepackage{mathrsfs}
\usepackage{tikz}
\usepackage{amsfonts}
\usepackage{kantlipsum}

\begin{document}

%\preprint{APS/123-QED}

\title{Large enhancement of the effective second-order nonlinearity in graphene metasurfaces}%

\author{Qun Ren}
%\email{qun.ren.15@ucl.ac.uk}
%\affiliation{ University College London}%

\author{J. W. You}%
%\email{j.you@ucl.ac.uk}
%\affiliation{ University College London}%

\author{N. C. Panoiu}
%\email{n.panoiu@ucl.ac.uk}
%\affiliation{ University College London}%

\affiliation{%
Department of Electronic and Electrical Engineering, University College London, Torrington Place, London WC1E 7JE, United Kingdom
}%

\date{\today}% It is always \today, today,

\begin{abstract}
Using a powerful homogenization technique, one- and two-dimensional graphene metasurfaces are
homogenized both at the fundamental frequency (FF) and second-harmonic (SH). In both cases, there
is excellent agreement between the predictions of the homogenization method and those based on
rigorous numerical solutions of Maxwell equations. The homogenization technique is then employed to
demonstrate that, owing to a double-resonant plasmon excitation mechanism that leads to strong,
simultaneous field enhancement at the FF and SH, the effective second-order susceptibility of
graphene metasurfaces can be enhanced by more than three orders of magnitude as compared to the
intrinsic second-order susceptibility of a graphene sheet placed on the same substrate. In
addition, we explore the implications of our results to the development of new active nanodevices
that incorporate nanopatterned graphene structures.
\end{abstract}

\maketitle

\section{Introduction}\label{Intro}

Metamaterials, which consist of artificial elements (so-called metaatoms or metamolecules) usually
arranged in a periodic pattern, have been playing an increasingly important r\^{o}le in
applications in which they emulate physical properties that otherwise cannot be achieved with
naturally occurring materials. The broad available choice of particular geometries and material
parameters of the constituents of metamaterials facilitates their use for the implementation of key
functionalities, including, \textit{inter alia}, phase engineering \cite{aman07prb,tcr14sr,yp18oe},
light focusing \cite{vgbc15josab, gzcql10oe, cz10oe}, and local field enhancement \cite{igmha13prb,
pegwc11jo, jwzxs14oe, baj09prb}. These functionalities are beginning to impact a series of research
fields by finding applications to bio-sensing \cite{apswg09nm, kymem16nm, cjxsn13acs, ngva15oe},
development of efficient absorbers \cite{nsjd08prl, rmcf12oe, sqlf13n}, electromagnetic cloaking
\cite{djbs06s, dyygs12nc}, and imaging beyond sub-diffraction limit
\cite{zshyycx07oe,dbcn10oe,ss12njp,gm15prl}. Among these physical properties of metamaterials,
local field enhancement is particularly relevant to nonlinear optics, as in this case the optical
response of a metamaterial-based device depends nonlinearly on the externally applied optical field
and thus can be widely tuned.

In many applications, the two-dimensional (2D) counterpart of metamaterials, the so-called
metasurfaces, can provide the required functionality, especially in the case of devices with planar
configuration. In addition, metasurfaces have the advantage of requiring much less laborious
fabrication processes. Moreover, in many applications pertaining to nonlinear optics, especially
those related to surface science and sensing, achieving the phase-matching of the interacting waves
is not a prerequisite condition, and therefore the constraints imposed on metasurfaces in order to
attain optimal energy conversion in nonlinear processes can be greatly relaxed \cite{jmjma15prb,
ad11ome, cga14prb, egy16nc,gst17nrm}.

Broadly speaking, there are two classes of optical metasurfaces: plasmonic metasurfaces based on
metallic particles \cite{fllfc16sa, ya11prb} and dielectric metasurfaces \cite{yidj14nc,
jimed15acs} relying on Mie resonances of dielectric particles. In the case of plasmonic
metasurfaces, the local field can be dramatically enhanced at plasmon-resonance frequencies
\cite{m07book,zmm05pr,vme10rmp,psl18jo}; however, this effect is usually accompanied by a
relatively large optical loss \cite{j15nn}. On the other hand, dielectric metasurfaces are
characterized by much smaller optical losses but usually provide reduced optical field enhancement.

A promising alternative to plasmonic and all-dielectric metasurfaces is provided by graphene
metasurfaces, as the (plasmon) resonance frequency of graphene nanostructures lies in the terahertz
domain, namely where optical losses of graphene are relatively small. Equally important, the
plasmonic nature of these resonances ensures that strong field enhancement can be achieved in
graphene metasurfaces, too. In addition, the corresponding resonance wavelength is much larger than
the size of graphene resonators, which means that a large number of such resonators can be packed
inside a domain with size comparable to that of the operating wavelength. Consequently, the optical
response of graphene metasurfaces can be highly isotropic, when the geometry of graphene unit cell
is symmetric. In fact, patterned graphene has already been employed in the design of terahertz
devices, such as perfect absorbers, filters, and tunable reflectors
\cite{tone16pre,toge18prb,vte18pra,fdf11nl,smt12nm,ybb19nanophot}. In this context, a particularly
appealing physical property of graphene is the tunability of its dielectric constant, a unique
functionality that is highly relevant to the design of active photonic devices.

In this paper, we propose a powerful and versatile homogenization approach for graphene
metasurfaces, and subsequently use it to demonstrate that the effective second-order susceptibility
of such metasurfaces can be dramatically increased due to the field-enhancement effect at plasmon
resonances. The novelty of the homogenization method used in this study consists in its ability to
describe not only metasurfaces containing linear and isotropic materials, like the standard
field-average methods, but also those made of anisotropic and nonlinear optical media. In addition,
we find that when a so-called double-resonance phenomenon occurs in a graphene metasurface
\cite{yymp17ptrsa}, the second-harmonic generation (SHG) can be further enhanced, leading to an
overall increase in SHG of more than three orders of magnitude as compared to the SHG of a graphene
sheet placed on the same substrate.

The paper is organized as follows: In the next section, the configurations of the graphene
metasurfaces investigated in this work are described, as well as their material parameters. In
Section~\ref{Theory}, an improved homogenization approach for retrieving the effective linear and
nonlinear properties of graphene metasurfaces is presented. Then, using this homogenization method,
the geometrical parameters of the graphene metasurfaces are optimized so as to achieve plasmon
resonances at both the fundamental frequency (FF) and second-harmonic (SH). In
Section~\ref{ResDisc}, the linear and nonlinear optical spectra of the graphene metasurfaces are
calculated and a comparison of the effective second-order susceptibility of graphene metasurfaces
with the second-order susceptibility of a graphene sheet placed on the same substrate is provided.
Finally, the main conclusions are outlined in Section~\ref{Concl}.

\section{Physical configuration and material parameters of graphene metasurfaces}\label{Conf}

\begin{figure}[!t]
\centering
\includegraphics[width=\columnwidth]{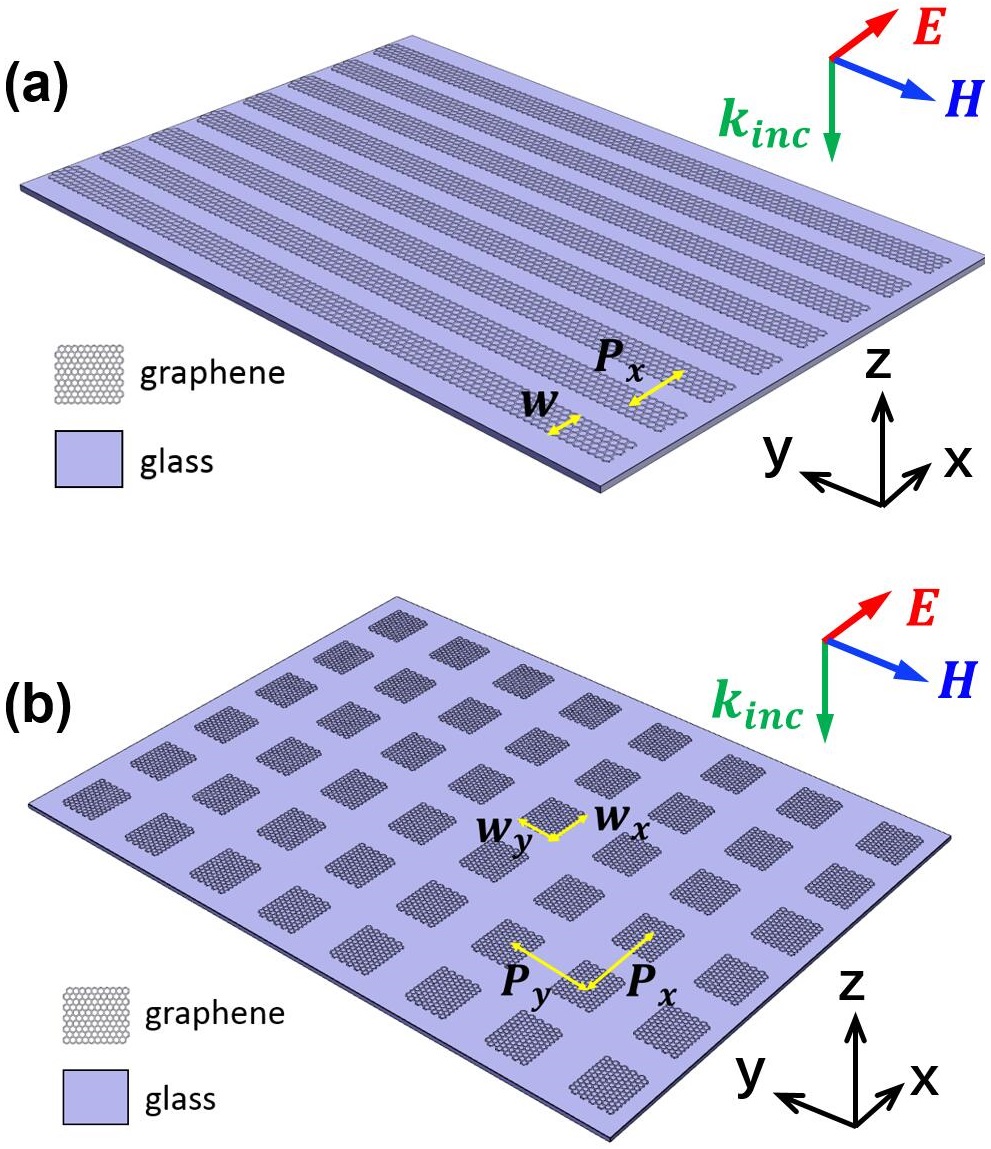}
\caption{(a) Schematics of a 1D graphene metasurface, with the period $P_{x}$ and width of graphene
ribbons, $w$. (b) Schematics of a 2D graphene metasurface, with periods $P_{x}$ and $P_{y}$, and
side-length of the graphene patches of $w_{x}$ and $w_{y}$. The two graphene metasurfaces are
illuminated by an $x$-polarized plane wave normally incident onto the metasurfaces.}
\label{schematic}
\end{figure}

In this section, we present the configuration of the one-dimensional (1D) and 2D graphene
metasurfaces studied in this work and describe the properties of the linear and nonlinear optical
constants of graphene. Thus, the two generic nonlinear graphene-based metasurfaces, a 1D
metasurface consisting of a periodic arrangement of graphene ribbons and a 2D metasurface
consisting of a rectangular array of graphene rectangular patches, are schematically illustrated in
Figs.~\ref{schematic}(a) and \ref{schematic}(b), respectively. The period of the 1D metasurface is
$P_{x}=\SI{100}{\nano\meter}$ and the width of the nanoribbons is $w$, whereas in the case of the
2D metasurface the periods along the $x$- and $y$-axis are $P_{x}=P_{y}=\SI{100}{\nano\meter}$ and
the length of the graphene patches along the $y$-axis is fixed at $w_{y}=\SI{30}{\nano\meter}$. The
width of the graphene nanoribbons and the length of the graphene patches along the $x$-axis,
$w_{x}$, are free parameters that will be optimized so as to achieve a double-resonance effect. In
both cases the graphene nanostructures are placed onto a silica substrate with
$n_{\mathrm{SiO_2}}=1.4$ and are illuminated by a normally incident, $x$-polarized plane wave with
field amplitude $E_{0}=\SI{1}{\volt\per\meter}$ (wave intensity
$I_{0}=\SI{4.43e12}{\watt\per\square\meter}$). This choice of the wave polarization ensures that
graphene plasmons exist in both metasurfaces.

Due to its metallic characteristics in the terahertz and infrared spectral regions, graphene
supports surface-plasmon polaritons (SPPs), which are collective oscillations of free electrons. In
the case of finite-size graphene nanostructures, the resonance frequency of SPPs is geometry
dependent. Therefore, by properly choosing the size and shape of these graphene nanostructures, one
can achieve a double-resonant phenomenon, namely SPPs exist both at the FF and SH. When this
occurs, the optical near-fields at the FF and SH are strongly enhanced, which leads to a marked
increase of the intensity of the SHG. Under these circumstances, one expects that the graphene
metasurface can be viewed as a homogeneous sheet of nonlinear material with strongly enhanced
effective second-order susceptibility.

Before we analyze in more detail the linear and nonlinear optical properties of the two graphene
metasurfaces, we briefly summarize the optical properties of the main optical constants of
graphene. Since graphene is a 2D semimetal, a surface optical conductivity, $\sigma_{s}$, is
generally used to describe its main linear physical properties at optical frequencies. Based on
Kubo's formula derived within the random-phase approximation, $\sigma_{s}$ can be expressed as the
sum of the intra-band ($\sigma_{\mathrm{intra}}$) and inter-band ($\sigma_{\mathrm{inter}}$)
contributions, $\sigma_{s}=\sigma_{\mathrm{intra}}+\sigma_{\mathrm{inter}}$. The intra-band part is
given by:
\begin{equation}
\label{eq:intra_band_conduct}
\sigma_{\mathrm{intra}}=\frac{e^{2}k_{B}T_{\tau}}{\pi\hbar^{2}(1-i\pi\tau)}\left[
\frac{\mu_{c}}{k_{B}T}+2\ln(e^{-\frac{\mu_{c}}{k_{B}T}}+1) \right],
\end{equation}
where $\mu_{c}$ is the chemical potential, $\tau$ is the relaxation time, $T$ is the temperature,
$e$ is the electron charge, $k_{B}$ is the Boltzmann constant, and $\hbar$ is the reduced Planck's
constant. Throughout our analysis, we use $\mu_{c}=\SI{0.6}{\electronvolt}$,
$\tau=\SI{0.25}{\pico\second}$, and $T=\SI{300}{\kelvin}$. Moreover, if $\mu_{c}\gg k_{B}T$, which
usually holds at room temperature, the inter-band part can be approximated as:
\begin{equation}
\label{eq:inter_band_conduct} \sigma_{\mathrm{inter}}= \frac{i e^{2}}{4\pi\hbar}\ln
\left[\frac{2|\mu_{c}|-(\omega+i\tau^{-1})\hbar}{2|\mu_{c}|+(\omega+i\tau^{-1})\hbar}\right].
\end{equation}
\begin{figure}[!t]
\centering
\includegraphics[width=\columnwidth]{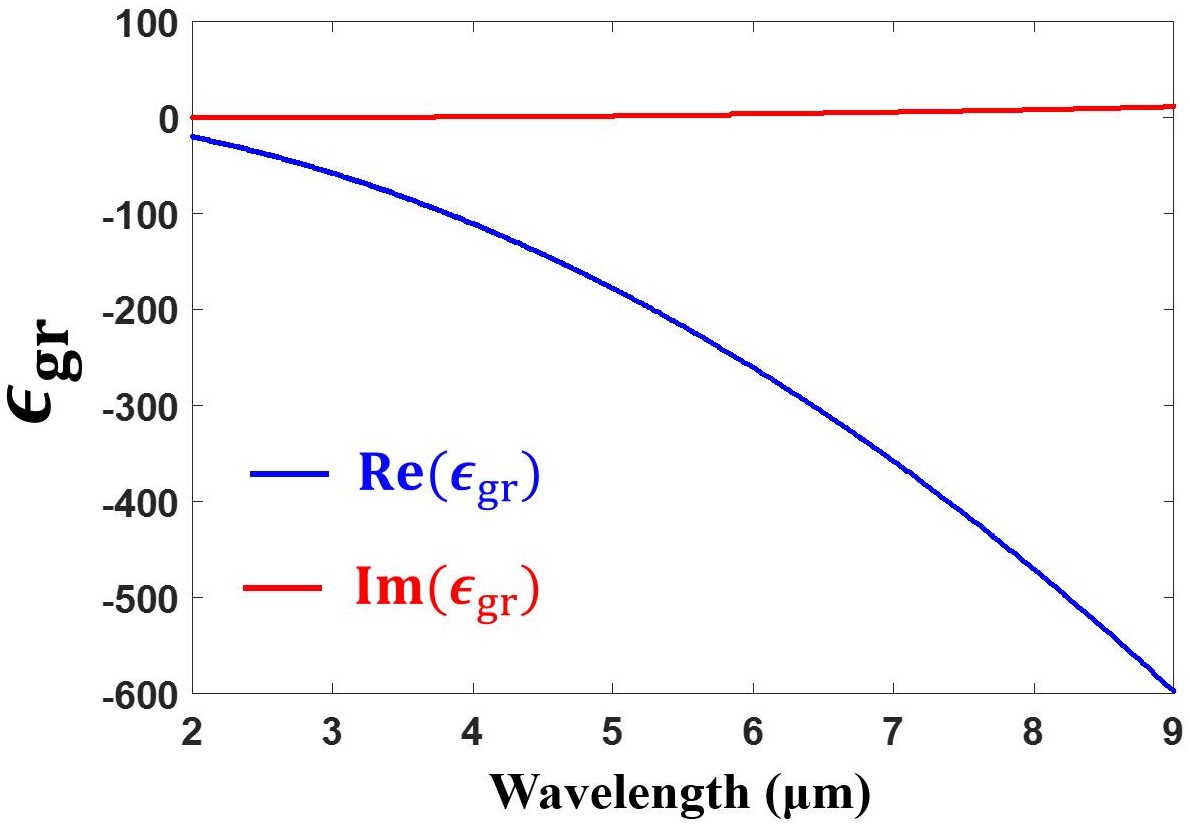}
\caption{Relative electric permittivity of a graphene sheet with
$h_{\mathrm{\mathrm{eff}}}=\SI{0.3}{\nano\meter}$.} \label{epsilon}
\end{figure}

If we assume that the effective thickness of graphene is $h_{\mathrm{\mathrm{eff}}}$, the relative
electric permittivity can be calculated from the conductivity through the relation:
\begin{equation}
\label{eq:relation_epsilon_conduct}
\epsilon_{\mathrm{gr}}(\omega)=1+\frac{i\sigma_{s}}{\epsilon_{0}\omega h_{\mathrm{eff}}}.
\end{equation}
The relative electric permittivity of graphene is depicted in Fig.~\ref{epsilon}, where
$h_{\mathrm{\mathrm{eff}}}=\SI{0.3}{\nano\meter}$ has been used.

Similar to the case of three-dimensional (3D) bulk optical media, the nonlinear optical properties
of 2D materials are generally determined by the symmetry properties of their atomic lattice and are
quantified by (bulk) nonlinear susceptibility tensors, $\bm{\chi}^{(n)}(\Omega;\omega)$, where
$\omega$ and $\Omega$ are the frequencies at the FF and higher-harmonic, respectively, and $n$ is
the order of the nonlinear optical process, or, equivalently, by surface nonlinear optical
conductivities, $\bm{\sigma}_{s}^{(n)}(\Omega;\omega)$. These two physical quantities are related
\textit{via} the following relation:
\begin{equation}\label{chisig}
    \bm{\chi}^{(n)}(\Omega;\omega)=\frac{i}{\epsilon_{0}\Omega
    h_{\mathrm{eff}}}\bm{\sigma}_{s}^{(n)}(\Omega;\omega).
\end{equation}

Free-standing graphene is a centrosymmetric material and therefore second-order nonlinear optical
processes and, in particular, SHG, are forbidden. If a graphene sheet, however, is placed onto a
homogeneous substrate the inversion symmetry is broken and (dipole) SHG is allowed. In particular,
such an optical configuration is characterized by a surface second-order nonlinear optical
conductivity tensor, $\bm{\sigma}_{s}^{(2)}(\Omega;\omega)$, where $\Omega=2\omega$. Symmetry
considerations based on the fact that graphene belongs to the $\mathcal{D}_{\mathrm{6h}}$ symmetry
group lead to the conclusion that this tensor has three independent nonzero components,
$\sigma_{s,\perp\perp\perp}^{(2)}$,
$\sigma_{s,\parallel\parallel\perp}^{(2)}=\sigma_{s,\parallel\perp\parallel}^{(2)}$, and
$\sigma_{s,\perp\parallel\parallel}^{(2)}$, where the symbols $\perp$ and $\parallel$ refer to the
directions perpendicular onto and parallel to the plane of graphene, respectively. The values of
these parameters used in this paper are
$\sigma_{s,\perp\perp\perp}^{(2)}=\SI[output-complex-root=\text{\ensuremath{i}}]{-9.71ie-16}{\ampere\meter\per\square\volt}$,
$\sigma_{s,\parallel\parallel\perp}^{(2)}=\sigma_{s,\parallel\perp\parallel}^{(2)}=\SI[output-complex-root=\text{\ensuremath{i}}]{-2.56ie-16}{\ampere\meter\per\square\volt}$,
and
$\sigma_{s,\perp\parallel\parallel}^{(2)}=\SI[output-complex-root=\text{\ensuremath{i}}]{-2.09ie-16}{\ampere\meter\per\square\volt}$,\cite{yjdja14prb,jm10prb}
and correspond to graphene placed on a silica substrate. Note that similar to the case of surface
nonlinear second-order susceptibility of nobel metals, the dominant component of the surface
nonlinear second-order conductivity (susceptibility) is the $\sigma_{s,\perp\perp\perp}^{(2)}$
($\chi_{s,\perp\perp\perp}^{(2)}$) component.

\section{Theory of linear and nonlinear homogenization}\label{Theory}

In this section, we describe a theoretical method we recently introduced \cite{jp19ol} for the
homogenization of the linear and nonlinear optical response of graphene metasurfaces. In
particular, we present an approach for extracting the effective linear and nonlinear optical
coefficients of a homogenized layer of material, which in the far-field has the same linear and
nonlinear optical response as that of the graphene metasurface. To be more specific, we use this
method to compute the effective electric permittivity of the two generic graphene metasurfaces, as
well as the effective surface second-order susceptibility of graphene metasurfaces, when they are
optimized to achieved maximum nonlinearity enhancement. Note that although the homogenized
metasurfaces can be characterized by effective surface quantities, such as linear and nonlinear
surface conductivities \cite{hkd11awpl,tks18acsp}, in this work we consider that the homogenized
metasurfaces have a finite thickness, $h_{\mathrm{eff}}$, and thus are described by bulk effective
permittivities and nonlinear susceptibilities.
\begin{figure}[!b]
\centering
\includegraphics[width=\columnwidth]{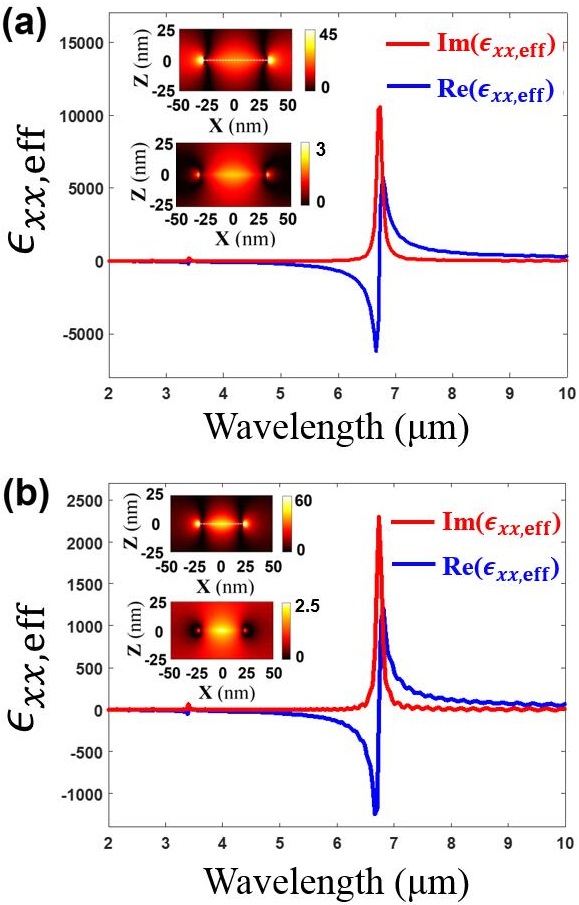}
\caption{(a) Effective relative permittivity of homogenized graphene-nanoribbon metasurface with
$w=\SI{57.5}{\nano\meter}$. In insets, the spatial profile of $\vert E_{x}\vert$, calculated at the
resonance wavelength $\lambda=\SI{6.74}{\micro\meter}$ (top panel) and at
$\lambda=\SI{4}{\micro\meter}$ (bottom panel). (b) The same as in (a), but calculated for the 2D
graphene metasurface with $w_{x}=\SI{42.5}{\nano\meter}$. The resonance wavelength for the 2D
graphene metasurface is $\lambda=\SI{6.93}{\micro\meter}$}\label{ribb_epsilon_eff}
\end{figure}

In order to develop a general homogenization method, we extend the traditional field-averaged
method to include nonlinear optical effects and anisotropic 2D materials. Thus, the constitutive
relation of a linear anisotropic material is expressed as:
\begin{equation}\label{const}
D_{i}=\sum_{j}\epsilon_{ij}E_{j},
\end{equation}
where $\mathbf{D}$ and $\mathbf{E}$ are the electric displacement and electric field, respectively,
and the subscripts $i,j = x,y,z$. Then, we introduce the averaged fields, defined as:
\begin{subequations}\label{avfields}
\begin{align}
\overline{\textbf{D}}_{\mathrm{eff}}(\omega)&=\frac{1}{V}\int_{V}\textbf{D}(\textbf{r},\omega)d\textbf{r},\label{Df} \\
\overline{\textbf{E}}_{\mathrm{eff}}(\omega)&=\frac{1}{V}\int_{V}\textbf{E}(\textbf{r},\omega)d\textbf{r},\label{Ef}
\end{align}
\end{subequations}
where $V$ is the volume of the unit cell of the (1D or 2D) metasurface. More specifically, the
integration domains for the 1D and 2D metasurfaces are $V=[0,P_{x}]\times[0,h_{\mathrm{eff}}]$ and
$V=[0,P_{x}]\times[0,P_{y}]\times[0,h_{\mathrm{eff}}]$, respectively. Using Eqs.~\eqref{const} and
\eqref{avfields}, the effective electric permittivity tensor of the metasurface, defined by the
constitutive relation
$\overline{D}_{i,\mathrm{eff}}=\sum_{j}\overline{\epsilon}_{ij,\mathrm{eff}}\overline{E}_{j,\mathrm{eff}}$,
can be written as:
\begin{equation}\label{epseff}
\overline{\bm{\epsilon}}_{ij,\mathrm{eff}}(\omega)=\frac{\displaystyle
\int_{V}\textbf{D}_{i}(\textbf{r},\omega)d\textbf{r}}{\displaystyle
\int_{V}\textbf{E}_{j}(\textbf{r},\omega)d\textbf{r}}= \frac{\displaystyle
\int_{V}\epsilon(\mathbf{r})\textbf{E}_{i}(\textbf{r},\omega)d\textbf{r}}{\displaystyle
\int_{V}\textbf{E}_{j}(\textbf{r},\omega)d\textbf{r}},
\end{equation}
where $\epsilon(\mathbf{r})=\epsilon_{0}$ if $\mathbf{r}$ is in air and
$\epsilon(\mathbf{r})=\epsilon_{\mathrm{gr}}$ if $\mathbf{r}$ is in graphene. The formula above has
been derived for metasurfaces made of isotropic optical materials, but it can be easily extended
to anisotropic ones.

In order to assess the validity of our homogenization method, we have calculated the effective
permittivity given by Eq.~\eqref{epseff} and then compared the optical response of the homogenized
metasurfaces, \textit{i.e.}, the absorption, $A$, transmittance, $T$, and reflectance, $R$, with
that of the two graphene metasurfaces. The optical near-fields needed to calculate
$\overline{\bm{\epsilon}}_{ij,\mathrm{eff}}(\omega)$, as well as the absorption, transmittance, and
reflectance of the two graphene metasurfaces, were computed using an in-house developed code
\cite{wp16prb,ytgp18josab}.
\begin{figure}[!t]
\centering
\includegraphics[width=\columnwidth]{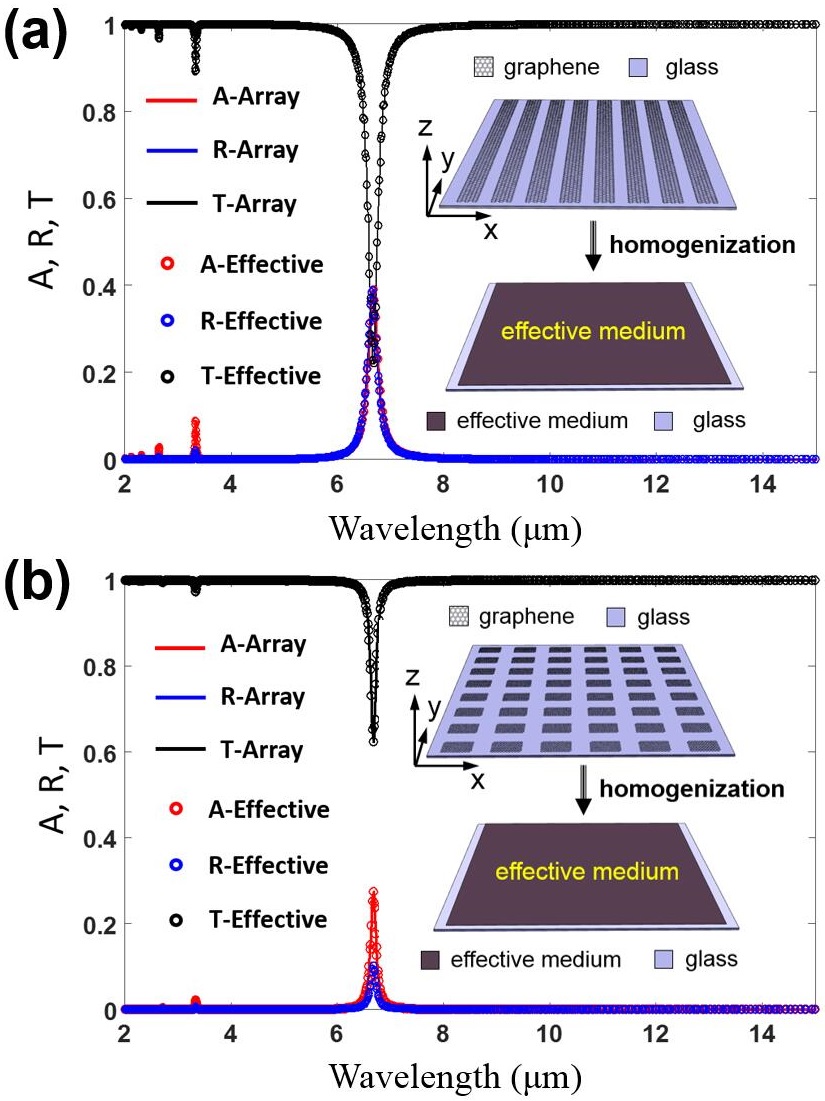}
\caption{Linear response comparison of absorption, $A$, reflectance, $R$, and transmittance, $T$,
calculated for the two graphene metasurfaces whose effective permittivities are presented in
Fig.~\ref{ribb_epsilon_eff} (depicted with solid curves) and $A$, $R$, and $T$ corresponding to
their homogenized counterparts (depicted with dotted curves).}\label{homo_test}
\end{figure}

The effective permittivities of the homogenized metasurfaces, $\epsilon_{xx,\mathrm{eff}}(\omega)$,
retrieved using the algorithm just described, are presented in Fig.~\ref{ribb_epsilon_eff}. The 1D
and 2D metasurfaces considered here were optimized for maximum nonlinear response using an approach
that will be described in the next section, the corresponding values of the geometrical parameters
being $w=\SI{57.5}{\nano\meter}$ and $w_{x}=\SI{42.5}{\nano\meter}$, respectively. In contrast to
the intrinsic permittivity of a homogeneous graphene sheet shown in Fig.~\ref{epsilon}, the
effective permittivities of the homogenized metasurfaces exhibit an evident Lorentzian resonant
response around a wavelength of about \SI{6.8}{\micro\meter}, which is reminiscent of the linear
optical response of an optical medium containing Lorentz-type resonators.

The field profiles presented in the insets of Figs.~\ref{ribb_epsilon_eff}(a) and
\ref{ribb_epsilon_eff}(b) suggest that at resonance the optical near-field is strongly enhanced,
which is one of the main physical properties of SPPs. Moreover, Fig.~\ref{ribb_epsilon_eff} shows
that in addition to this main resonance, few other higher-order resonances exist at smaller
wavelengths. These higher-order resonances correspond to the excitation of higher-order plasmon
modes in the graphene nanoribbons or graphene patches. Interestingly enough, although graphene has
metallic characteristics in the frequency range considered in our calculations, near the resonance
$\mathfrak{Re}(\overline{\bm{\epsilon}}_{xx,\mathrm{eff}})>0$, which means that the homogenized
metasurfaces behave as a dielectric around this frequency.

The main aim of a homogenization theory is to reduce a patterned metasurface to a homogeneous sheet
characterized by certain effective optical constants. A reliable way to assess the validity of this
procedure is to compare the optical response of the homogenized metasurface and the original one,
as quantified by physical quantities such as absorption, reflectance, and transmittance. We
performed this analysis for the two graphene metasurfaces whose effective permittivities are
presented in Fig.~\ref{ribb_epsilon_eff}, the corresponding results being summarized in
Fig.~\ref{homo_test}. This comparison clearly demonstrates that the linear response of the
homogenized sheets perfectly agrees with that of the original graphene metasurfaces, thus proving
the accuracy of the proposed linear homogenization approach. This is explained by the fact that the
wavelengths considered in our computations, including those at which the graphene metasurfaces are
strongly resonant, are much larger than the characteristic size of the graphene constituents of the
metasurfaces, so that the two optical structures are operated deep in the metasurface regime.

We now extend the homogenization method to the nonlinear regime, and use SHG as an illustrative
nonlinear optical process. Thus, this nonlinear optical interaction is determined by the following
nonlinear polarization:
\begin{equation}
\label{eq:Polarization}
\textbf{P}(\Omega;\textbf{r})=\epsilon_{0}\bm{\chi}^{(2)}(\Omega;\textbf{r}):
\mathbf{E}(\omega;\textbf{r})\mathbf{E}(\omega;\textbf{r}),
\end{equation}
where $\Omega=2\omega$ and
$\bm{\chi}^{(2)}(\Omega;\textbf{r})=\bm{\chi}_{\mathrm{gr}}^{(2)}(\Omega)$ if $\mathbf{r}$ is in
graphene and $\bm{\chi}^{(2)}(\Omega;\textbf{r})=0$ if $\mathbf{r}$ is in air. Based on
Eq.~\eqref{eq:Polarization}, the components of the SH polarization can be evaluated as:
\begin{equation}\label{eq:Polarization_i}
P_{i}=\epsilon_{0}\sum_{jk}\chi_{ijk}^{(2)}E_{j}E_{k}\equiv\sum_{jk}q_{ijk},
\end{equation}
where we have introduced the auxiliary quantities,
$q_{ijk}=\epsilon_{0}\chi_{ijk}^{(2)}E_{j}E_{k}$. The averaged value of these auxiliary quantities
are:
\begin{equation}\label{eq:q_ijk}
\overline{q}_{ijk}(\Omega)=\frac{1}{V}\int\chi_{ijk}^{(2)}(\Omega;\textbf{r})
E_{j}(\omega;\textbf{r})E_{k}(\omega;\textbf{r})d\textbf{r}.
\end{equation}
\begin{figure}[!b]
\centering
\includegraphics[width=\columnwidth]{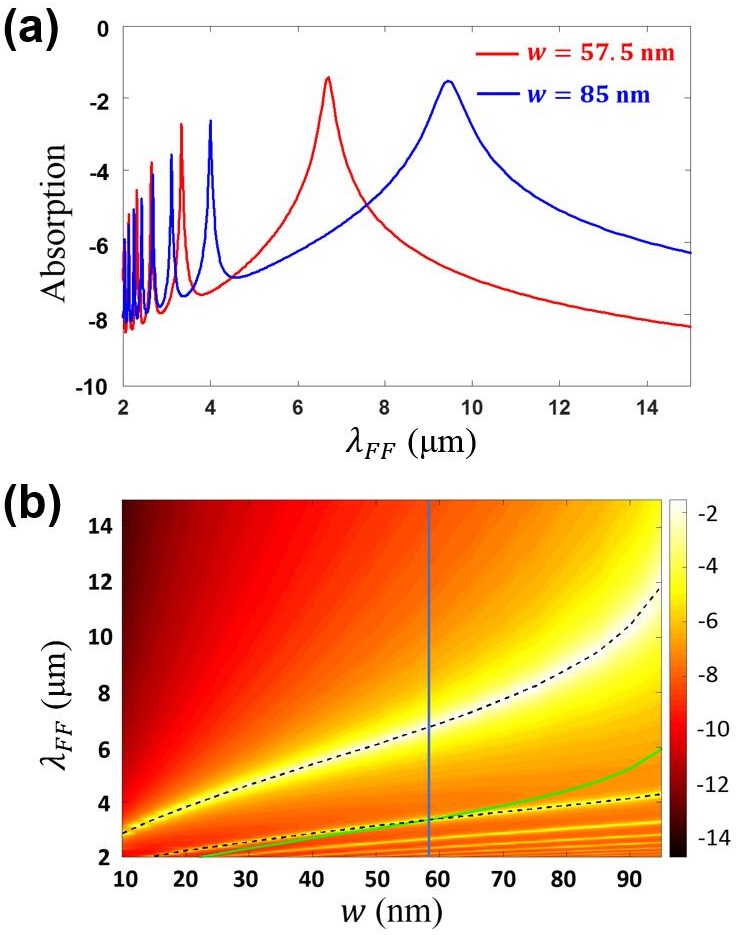}
\caption{(a) Absorption spectra of the 1D graphene metasurface presented in
Fig.~\ref{schematic}(a), calculated for the optimum width, $w=\SI{57.5}{\nano\meter}$, for which a
double-resonance phenomenon occurs, and for $w=\SI{85}{\nano\meter}$. (b) Dispersion map of
absorption. Dashed curves indicate the plasmon bands, whereas the green curve indicates the
half-wavelength of the fundamental plasmon band. The vertical line indicates that there is a
double-resonance effect for $w=\SI{57.5}{\nano\meter}$.}\label{FF_spec_ribb}
\end{figure}

Similarly to Eq.~\eqref{eq:Polarization}, the nonlinear SH polarization in the homogenized
metasurfaces can be written as:
\begin{equation}
\label{eq:Polarizationeff}
\overline{\textbf{P}}_{\mathrm{eff}}(\Omega)=\epsilon_{0}\bm{\chi}_{\mathrm{eff}}^{(2)}(\Omega):
\overline{\mathbf{E}}_{\mathrm{eff}}(\omega)\overline{\mathbf{E}}_{\mathrm{eff}}(\omega),
\end{equation}
where $\bm{\chi}_{\mathrm{eff}}^{(2)}(\Omega)$ is the effective second-order susceptibility of the
homogenized metasurface.

The homogenized metasurface and the original one will have the same nonlinear optical response in
the far-field if the averaged nonlinear polarization in Eq.~\eqref{eq:Polarization} is
\textit{termwise equal} to the effective nonlinear polarization described by
Eq.~\eqref{eq:Polarizationeff}. Using this condition, the effective second-order susceptibility of
the homogenized metasurface can be evaluated as:
\begin{equation}\label{eq:chi_eff}
\bm{\chi}_{\mathrm{eff},ijk}^{(2)}(\Omega)= \frac{\displaystyle
\overline{q}_{ijk}(\Omega)}{\displaystyle \overline{E}_{\mathrm{eff},j}(\omega)
\overline{E}_{\mathrm{eff},k}(\omega)}.
\end{equation}

\section{Results and Discussion}\label{ResDisc}

In this section, we describe our approach to optimize the nonlinear optical response of graphene
metasurfaces and quantify the nonlinearity enhancement of the optimized metasurfaces. In
particular, we calculate the effective second-order susceptibility of the graphene metasurfaces and
compare it to the second-order susceptibility of a graphene sheet placed onto the same silica
substrate.
\begin{figure}[!b]
\centering
\includegraphics[width=\columnwidth]{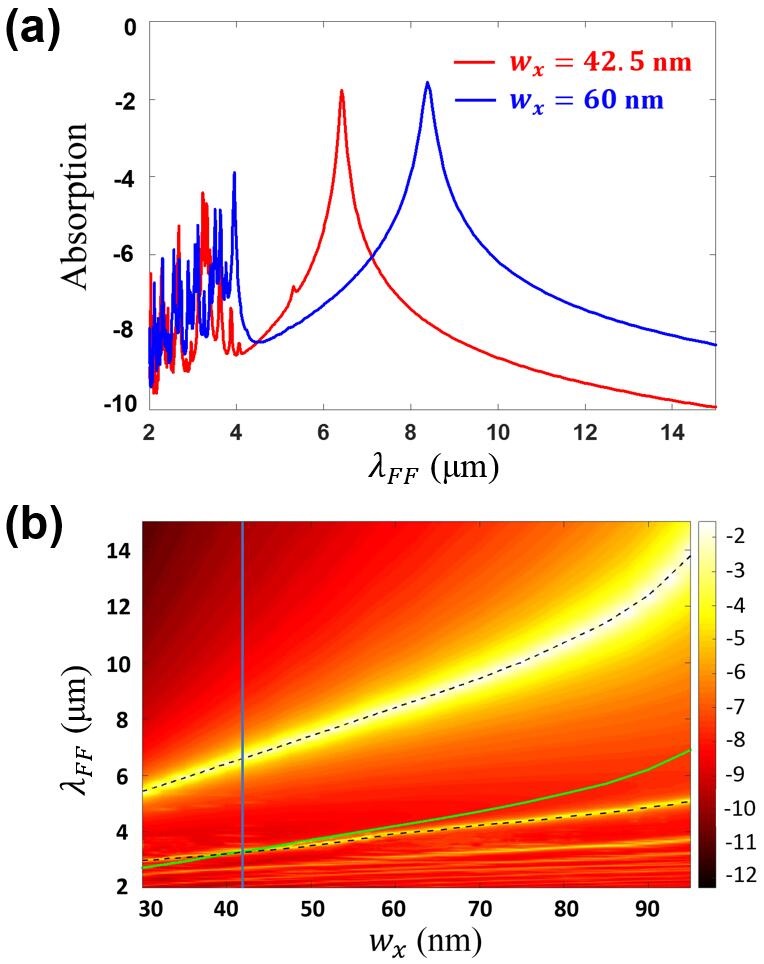}
\caption{(a) Absorption spectra of the 2D graphene metasurface presented in
Fig.~\ref{schematic}(b), calculated for the optimum side-length, $w_{x}=\SI{42.5}{\nano\meter}$,
for which a double-resonance phenomenon occurs, and for $w_{x}=\SI{60}{\nano\meter}$. (b)
Dispersion map of absorption. Dashed curves indicate the plasmon bands, whereas the green curve
indicates the half-wavelength of the fundamental plasmon band. The vertical line shows that there
is a double-resonance effect for $w_{x}=\SI{42.5}{\nano\meter}$.}\label{FF_spec_rect_x}
\end{figure}

\subsection{Linear optical response of 1D and 2D graphene metasurfaces}

One effective approach to achieve a significant enhancement of the SHG in graphene metasurfaces is
to engineer their geometrical parameters so as plasmons exist at both the FF and SH. Under these
conditions, the incoming light would in-couple effectively into the metasurface, as plasmons exist
at the FF, which would lead to a strong enhancement of the optical near-field at the FF, and, as
per Eq.~\eqref{eq:Polarization}, of the nonlinear polarization. Moreover, if plasmons exist at the
SH, too, the nonlinear sources will radiate efficiently into the continuum, the graphene
metasurface behaving in these conditions as an efficient nanoantenna.

One particularly useful tool for optimizing the linear and nonlinear optical response of graphene
metasurfaces is the dispersion map of the absorption, namely the dependence of the optical
absorption spectra on a certain parameter. Because the optical absorption increases when plasmons
are excited in the structure, the absorption dispersion map provides valuable information about the
frequency dispersion of the plasmon modes. The corresponding absorption spectra have been
calculated using a computational method \cite{wp16prb,ytgp18josab} that rigorously incorporates
both the frequency dispersion and nonlinearity of graphene.

We begin our analysis with the 1D graphene metasurface presented in Fig.~\ref{schematic}(a). Thus,
we show in Fig.~\ref{FF_spec_ribb}(a) the linear absorption spectra determined for the optimum
width of the graphene nanoribbons, $w=\SI{57.5}{\nano\meter}$ (we will explain letter how this
value was determined) and for some other arbitrary value, $w=\SI{85}{\nano\meter}$. Moreover, the
dispersion map of the optical absorption corresponding to this metasurface is plotted in
Fig.~\ref{FF_spec_ribb}(b). It can be seen in Fig.~\ref{FF_spec_ribb}(a) that the absorption
spectra present a series of plasmon resonances, whose amplitude decreases as the resonance
wavelength decreases. These resonances appear in the absorption map as a series of
geometry-dependent plasmon bands, indicated with dashed curves, with the resonance wavelength
increasing with the increase of the width of the nanoribbons. Importantly,
Fig.~\ref{FF_spec_ribb}(b) suggests that for $w=\SI{57.5}{\nano\meter}$ the nanoribbons support a
(fundamental) plasmon at the FF and a second-order plasmon at the SH, namely the metasurface
possesses a double-resonance feature.
\begin{figure}[!b]
\centering
\includegraphics[width=\columnwidth]{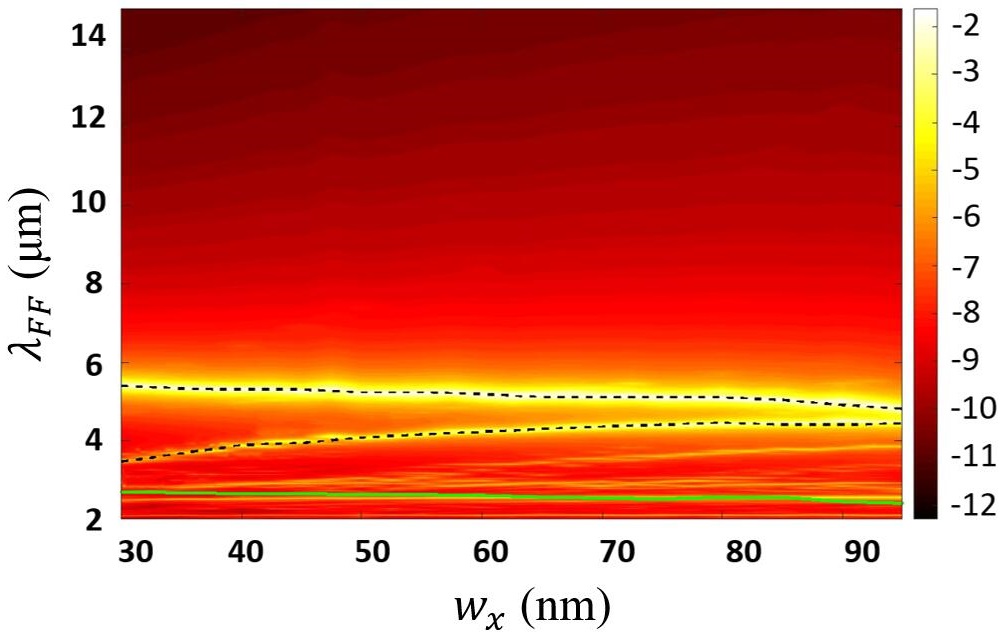}
\caption{The same as in Fig.~\ref{FF_spec_rect_x}(b) but determined for an $y$-polarized incident
plane wave.}\label{FF_spec_y}
\end{figure}

Similar conclusions can be drawn in the case of the 2D graphene metasurface. Thus, similarly to the
data summarized in Fig.~\ref{FF_spec_ribb}, we present in Figs.~\ref{FF_spec_rect_x}(a) and
\ref{FF_spec_rect_x}(b) two linear absorption spectra determined for the optimum side-length of the
graphene patches, $w_x=\SI{42.5}{\nano\meter}$, and for an arbitrary value,
$w_{x}=\SI{60}{\nano\meter}$, as well as the corresponding dispersion map of the optical
absorption, respectively. It can be seen that in the 2D case, too, the resonance wavelength of the
plasmon bands increases with $w_{x}$ and that the double-resonance phenomenon also occurs in 2D
graphene metasurfaces. To be more specific, if $w_x=\SI{42.5}{\nano\meter}$ plasmon resonances exit
at both the FF of $\lambda_{\mathrm{FF}}=\SI{6.93}{\micro\meter}$, which is a fundamental plasmon,
and at the SH of $\lambda_{\mathrm{SH}}=\lambda_{\mathrm{FF}}/2=\SI{3.47}{\micro\meter}$. Note
that, as illustrated in Fig.~\ref{schematic}(b), the 2D graphene metasurface is normally
illuminated by an $x$-polarized plane wave.

The 2D graphene metasurface is anisotropic and therefor the optical absorption spectra depend on
the polarization of the incident light. This idea is validated by the dispersion map of the optical
absorption shown in Fig.~\ref{FF_spec_y}, which has been determined for a normally incident,
$y$-polarized incident plane wave. Thus, for this wave polarization the wavelength of
fundamental-plasmon band increases with $w_{x}$, whereas the wavelength of the higher-order plasmon
bands decrease with $w_{x}$.

It can also be seen that when $w_{x}$ varies, the plasmon bands are more dispersive for
$x$-polarized incident waves as compared to those in the case of $y$-polarized waves. This finding
is explained by the fact that the wavelength of the plasmon resonance is primarily determined by
the size of the patch along the direction of the electric field. More importantly, however, the
results in Fig.~\ref{FF_spec_y} suggest that the double-resonance effect does not occur for
$y$-polarized incident plane waves. In our analysis, we have only considered $x$- and $y$-polarized
incident plane waves, chiefly because the conclusions for other polarizations can be derived from
the results corresponding to the linear superposition of these two primary polarizations.
\begin{figure}[!t]
\centering\includegraphics[width=\columnwidth]{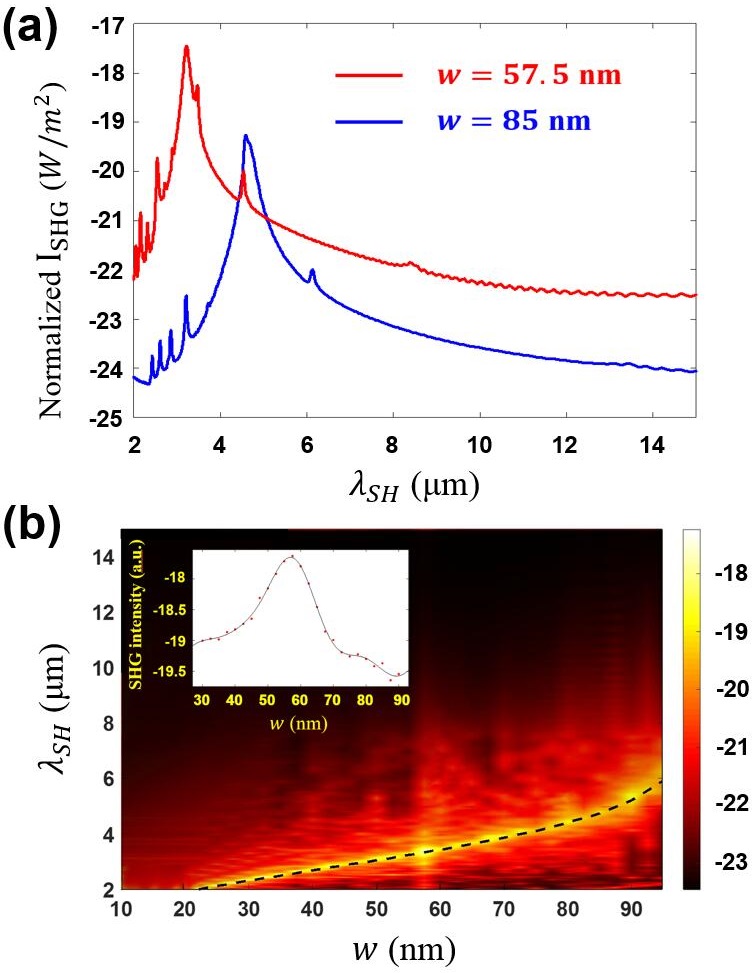} \caption{(a) Normalized SHG intensity
spectra, $I_{\mathrm{SHG}}$, of the 1D graphene metasurface presented in Fig.~\ref{schematic}(a),
calculated for the optimum width, $w=\SI{57.5}{\nano\meter}$, and for $w=\SI{85}{\nano\meter}$. (b)
Dispersion map of $I_{\mathrm{SHG}}$. The dashed curve indicates the fundamental-plasmon band. The
inset shows the dependence of $I_{\mathrm{SHG}}$ \textit{vs.} $w$, determined for the case when the
wavelengths of the FF and fundamental plasmon are the same.}\label{SH_spec_ribb}
\end{figure}

\begin{figure}[!t]
\centering
\includegraphics[width=\columnwidth]{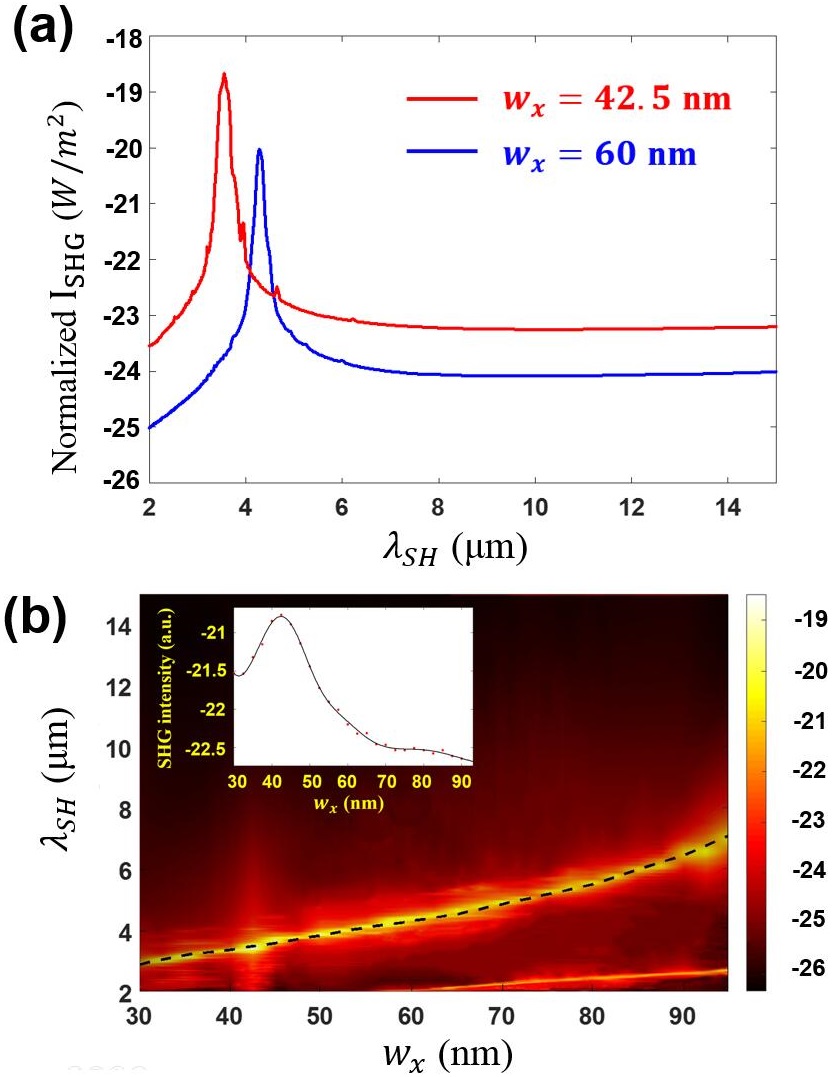}
\caption{(a) Normalized SHG intensity spectra, $I_{\mathrm{SHG}}$, of the 2D graphene metasurface
presented in Fig.~\ref{schematic}(b), calculated for the optimum side-length,
$w_{x}=\SI{42.5}{\nano\meter}$ and for $w_{x}=\SI{60}{\nano\meter}$. (b) Dispersion map of
$I_{\mathrm{SHG}}$. Dashed curve indicates the fundamental-plasmon band. The inset shows the
variation of $I_{\mathrm{SHG}}$ with $w_{x}$, computed for the case when the wavelengths of the FF
and fundamental plasmon are the same.}\label{SH_spec_rect}
\end{figure}

\subsection{Nonlinear optical response of 1D and 2D graphene metasurfaces}

We now turn our attention to SHG in 1D and 2D graphene metasurfaces and investigate the influence
of plasmon excitation at the FF and SH on the nonlinear optical response of the two graphene
metasurfaces. To this end, we used a generalized-source FDTD numerical method \cite{ytgp18josab},
to rigorously compute the SHG in the graphene metasurfaces. Since we want to compare the SHG
intensity corresponding to different values of the width of the nanoribbons and rectangular
patches, we normalize the SHG intensity to the area of the graphene structure contained in a unit
cell (note that the periods $P_{x}$ and $P_{y}$ are not changed, so that the area of the unit cells
do not vary). More specifically, the normalized SHG intensity spectra, $I_{\mathrm{SHG}}$, were
calculated as follows: in the 1D case we computed the SHG power per unit length and then divided
the result by the corresponding area of the graphene nanoribbon. In the 2D case, we computed the
SHG power coresponding to the unit cell with area $P_{x}\times P_{y}$ and divided the result to the
area of the graphene patch, $w_{x}\times w_{y}$. Note that the normalized SHG intensity represents
the sum of the SHG signals emitted in the transmission and reflection directions.

The results of these calculations are presented in Fig.~\ref{SH_spec_ribb} and
Fig.~\ref{SH_spec_rect} and correspond to the 1D and 2D metasurfaces, respectively. As
Eq.~\eqref{eq:Polarization} shows, the nonlinear polarization is proportional to the square of the
optical near-field at the FF and therefore the SHG intensity is proportional to the FF field
amplitude to the fourth. As a result, the resonance peaks of normalized SHG intensity spectra and
the plasmon bands of the corresponding dispersion maps of the normalized SHG intensity should be
observed at exactly the half-wavelength of the resonance peaks of linear optical absorption spectra
and the corresponding plasmon bands of the dispersion maps of the linear optical absorption. This
prediction is fully validated by a comparison between the results presented in
Fig.~\ref{FF_spec_ribb} and Fig.~\ref{SH_spec_ribb} on the one hand, results that correspond to the
1D graphene metasurface, and, on the other hand, the results plotted in Fig.~\ref{FF_spec_rect_x}
and Fig.~\ref{SH_spec_rect}, which correspond to the 2D graphene metasurface.
\begin{figure}[!b]
\centering
\includegraphics[width=\columnwidth]{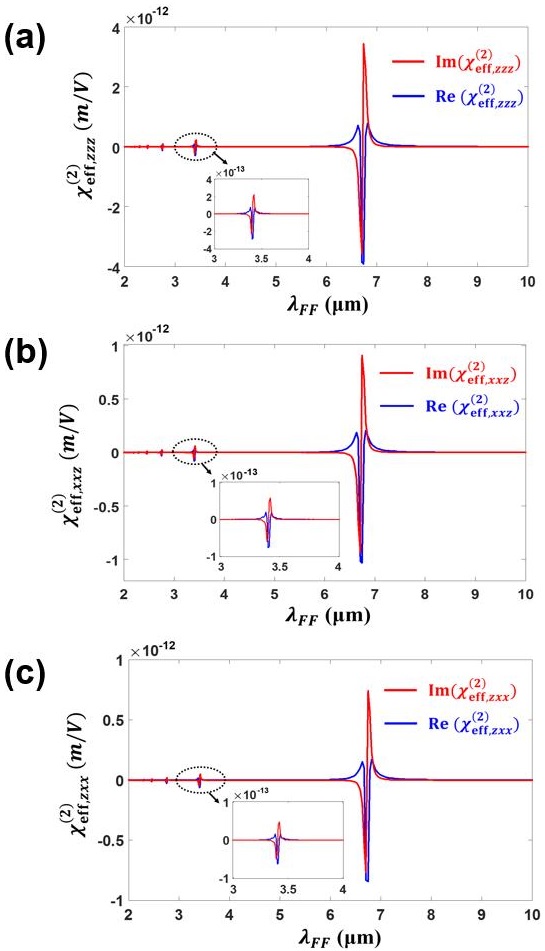}
\caption{Wavelength dependence of the three independent components of the effective second-order
susceptibility, $\bm{\chi}_{\mathrm{eff}}^{(2)}$, of the 1D graphene metasurface.
}\label{susceptiblity_333}
\end{figure}

Importantly, the insets in Fig.~\ref{SH_spec_ribb}(b) and Fig.~\ref{SH_spec_rect}(b) demonstrate
the SHG enhancement due to the double-resonance mechanism. Indeed, it can be inferred from these
plots that for the 1D graphene metasurface maximum SHG intensity is achieved for a width of the
graphene nanoribbons of $w=\SI{57.5}{\nano\meter}$, whereas in the case of the 2D graphene
metasurface the optimum value of the side-length of the graphene patch that leads to maximum SHG
intensity is $w_{x}=\SI{42.5}{\nano\meter}$. This clearly proves that in addition to
plasmon-enhanced SHG, the double-resonance mechanism can be employed to achieve further significant
enhancement of the nonlinear optical response of graphene metasurfaces.

\subsection{Enhancement of the effective second-harmonic susceptibility of 1D and 2D graphene metasurfaces}

\begin{figure}[!b]
\centering
\includegraphics[width=\columnwidth]{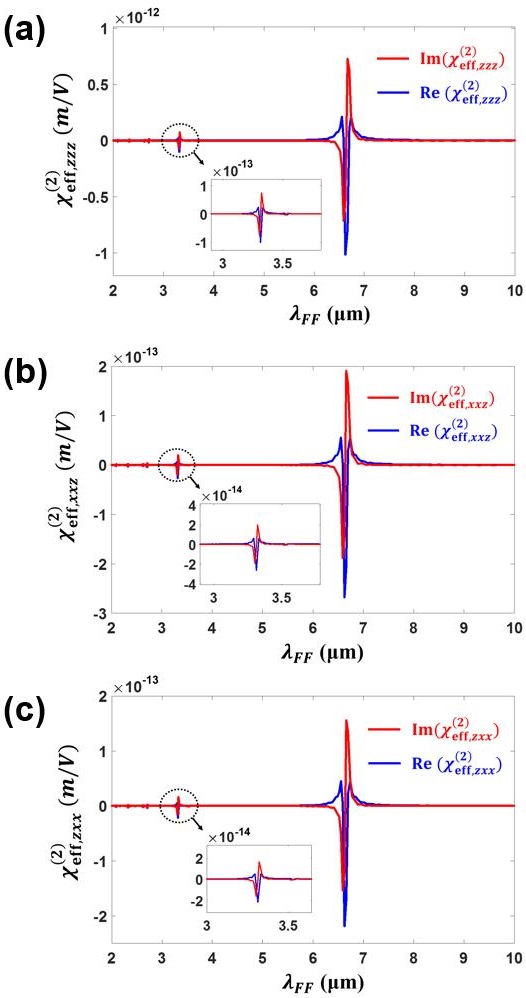}
\caption{The same as in Fig.~\ref{susceptiblity_333}, but corresponding to the 2D graphene
metasurface.}\label{susceptiblity_333_2}
\end{figure}

A suitable physical quantity that measures the enhancement of the nonlinear optical response of a
nonlinear optical system is the nonlinear susceptibility. Therefore, we have used the
homogenization method described in Sec.~\ref{Theory} to calculate the effective second-order
susceptibility of the two graphene metasurfaces. In particular, we retrieved the three independent
components of this nonlinear susceptibility, $\chi_{\mathrm{eff},zzz}^{(2)}$,
$\chi_{\mathrm{eff},xxz}^{(2)}$, and $\chi_{\mathrm{eff},zxx}^{(2)}$. The results of these
calculations are summarized in Fig.~\ref{susceptiblity_333} and Fig.~\ref{susceptiblity_333_2}, and
correspond to the 1D and 2D metasurfaces, respectively.

One important conclusion that can be inferred from the data presented in these figures is that,
similar to the case of the effective permittivity of the homogenized graphene metasurfaces, all
components of the effective second-order susceptibilities show a resonant behavior around the
plasmon resonance wavelength (fundamental and higher-order wavelength), which means that the
enhancement of the nonlinearity of the graphene metasurfaces can be traced to the excitation of
graphene SPPs. The maximum enhancement occurs when the fundamental plasmon is excited. Moreover,
the spectra of these components of the second-order susceptibilities are similar to those of a
nonlinear optical medium containing resonators of Lorentzian nature, which suggests that the
graphene nanostructures that constitute the building blocks of the two metasurfaces can be viewed
as metaatoms responsible for the effective nonlinear optical response of these optical
nanostructures. Since the size of these metaatoms is much smaller than the resonance wavelength at
the SH, one can conclude that the nonlinear graphene gratings investigated in this study operate in
the metasurface regime, too.
\begin{figure}[!t]
\centering
\includegraphics[width=0.9\columnwidth]{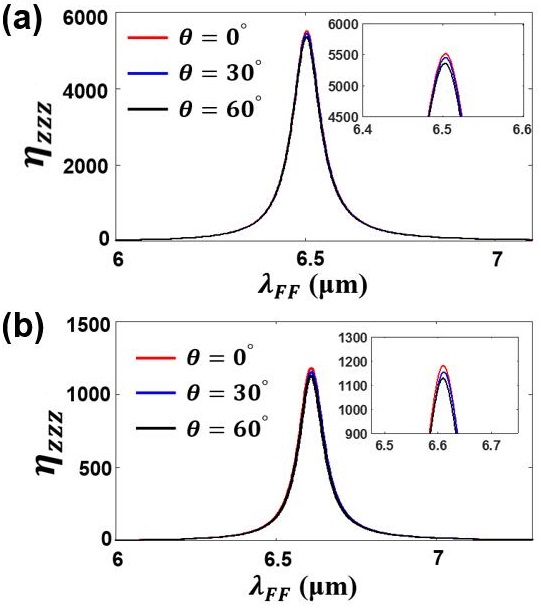}
\caption{(a) Wavelength dependence of the enhancement factor of the dominant component of the
effective second-order susceptibility, determined for the optimized 1D graphene metasurface for
several values of the angle of incidence, $\theta$. (b) The same as in (a), but determined for the
2D optimized graphene metasurface.}\label{enhancement}
\end{figure}

In order to further analyze the characteristics of the magnitude of the enhancement of the
nonlinear optical response of the two graphene metasurfaces, we also calculated the enhancement
factors $\eta_{zzz}=\vert\chi_{\mathrm{eff},zzz}^{(2)}/\chi_{\mathrm{gr},zzz}^{(2)}\vert$,
$\eta_{xxz}=\vert\chi_{\mathrm{eff},xxz}^{(2)}/\chi_{\mathrm{gr},xxz}^{(2)}\vert$, and
$\eta_{zxx}=\vert\chi_{\mathrm{eff},zxx}^{(2)}/\chi_{\mathrm{gr},zxx}^{(2)}\vert$ for several
different values of the angle of incidence, $\theta$. In these definitions,
$\bm{\chi}_{\mathrm{gr}}^{(2)}$ is the surface second-order susceptibility of a uniform graphene
sheet placed on top of a silica substrate.

The results of this analysis are summarized in Fig.~\ref{enhancement}, where we show the data
corresponding to the enhancement $\eta_{zzz}$ of the dominant component of
$\bm{\chi}_{\mathrm{eff}}^{(2)}$ of the 1D and 2D metasurfaces. This figure demonstrates a
remarkable enhancement of the second-order nonlinearity of the two metasurfaces, especially near
the plasmon resonance. In particular, the dominant component $\chi_{\mathrm{eff},zzz}^{(2)}$ of the
homogenized graphene metasurfaces is larger by more than three orders of magnitude than the
corresponding component $\chi_{\mathrm{gr},zzz}^{(2)}$ of a graphene sheet placed on the same
silica substrate. It can also be observed that $\eta_{zzz}$ only slightly decreases as the angle of
incidence increases, which further proves that the graphene elements of the metasurfaces behave as
true metaatoms. We also stress that despite the fact that the nonlinear optical losses are enhanced
as well around plasmon resonances, one expects that this is not a particularly detrimental effect
as the graphene metasurfaces investigated in this work are not meant to be employed in applications
where large propagation distances are required.

\section{Conclusion}\label{Concl}

In summary, in this study we investigated the optical response of one- and two-dimensional graphene
metasurfaces and their homogenized counterparts. In particular, using a recently developed
homogenization technique, we retrieved the effective permittivity and effective second-order
susceptibility of the homogenized metasurfaces and compared the values of several physical
quantities characterizing the original and homogenized metasurfaces, such as the optical
absorption, transmittance, and reflectance. Our analysis revealed that for metasurfaces whose
graphene constituents have characteristic size of a few tens of nanometers there is an excellent
agreement between the predictions of the homogenization method and the results obtained by
rigorously solving the Maxwell equations. This was explained by the fact that the characteristic
size of graphene resonators is much smaller than their resonance wavelength.

Our theoretical analysis of the two types of homogenized graphene metasurfaces showed that their
nonlinear response can be greatly enhanced when surface plasmons are excited in their graphene
constituents. Additional nonlinearity enhancement is achieved when plasmons exist at both the
fundamental-frequency and second-harmonic, the overall effect of this double-resonance effect being
an enhancement of the effective second-order susceptibility of the graphene metasurfaces by more
than three orders of magnitude. Moreover, it should be noted that this double-resonance phenomenon
could also be observed in other more complex configurations, \textit{e.g.} when plasmons are
excited in different plasmonic materials, such as metasurfaces containing coupled metallic-graphene
nanostructures. Equally important, the proposed homogenization method can be readily extended to
other cases, too, such as three-dimensional configurations or incident waves with arbitrary
polarization and angle of incidence, which further underscores the importance of the results
reported in this study.

\section*{Acknowledgments}\label{Ack}
This work was supported by the European Research Council (ERC), Grant Agreement No.
ERC-2014-CoG-648328, China Scholarship Council (CSC), and University College London (UCL)
(201506250086). The authors acknowledge the use of the UCL Legion High Performance Computing
Facility (Legion@UCL) and associated support services in the completion of this work.

\end{document}